\documentclass[aps,prl,floats,letterpaper,floatfix,nofootinbib,superscriptaddress,preprintnumbers,twocolumn]{revtex4}
\pdfoutput=1
\usepackage{amsmath}
\usepackage{color}
\usepackage[mathenv]{}
\usepackage{graphicx}
\usepackage{microtype}

\begin{document}

\title{A route to observing ponderomotive entanglement with optically
  trapped mirrors}

\author{Christopher Wipf}
\affiliation{LIGO Laboratory, Massachusetts Institute of Technology,
  Cambridge, MA 02139}
\author{Thomas Corbitt}
\affiliation{LIGO Laboratory, Massachusetts Institute of Technology,
  Cambridge, MA 02139}
\author{Yanbei Chen}
\affiliation{Theoretical Astrophysics, California Institute of
  Technology, Pasadena, CA 91125}
\author{Nergis Mavalvala}
\affiliation{LIGO Laboratory, Massachusetts Institute of Technology,
  Cambridge, MA 02139}

\begin{abstract}
  The radiation pressure of two detuned laser beams can create a
  stable trap for a suspended cavity mirror; here it is shown that
  such a configuration entangles the output light fields via
  interaction with the mirror.  Intra-cavity, the opto-mechanical
  system can become entangled also.  The degree of entanglement is
  quantified spectrally using the logarithmic negativity.
  Entanglement survives in the experimentally accessible regime of
  gram-scale masses subject to thermal noise at room temperature.
\end{abstract}

%\pacs{}

\definecolor{purple}{rgb}{0.6,0,1}
\preprint{\large \color{purple} LIGO-P070064-00-Z}
\date{\today}
\maketitle

Entanglement both provides a basis for fundamental tests of quantum
mechanics, and is an ingredient for applications in quantum
information, including cryptography and teleportation.  Producing
entanglement in a macroscopic mechanical system has become a prominent
experimental objective, and progress in the fabrication and cooling of
small mechanical resonators is quickly bringing this objective within
reach~\cite{2004Natur.432.1002M, 2006Natur.443..193N,
  2006Natur.444...67G, 2006Natur.444...71A, 2006Natur.444...75K,
  schliesser:243905, harris:013107, poggio:017201}, as highlighted in
a series of recent proposals treating these
systems~\cite{PhysRevA.56.4175, PhysRevLett.88.120401,
  PhysRevLett.88.148301, PhysRevA.68.013808, PhysRevLett.91.130401,
  2003qic-gmt-information-theoretic-ponderomotive, ferreira:060407,
  PhysRevLett.97.150403, PhysRevLett.98.030405, paternostro:250401}.

Meanwhile, the improving sensitivity of gravitational-wave
interferometers is opening a new regime for macroscopic quantum
mechanics, and may reveal quantum features such as squeezing and
entanglement of their mirrors' motion~\cite{mueller-ebhardt:013601}.
A novel and defining property of this regime is that radiation
pressure effects, in particular the optical spring~\cite{
  2002PhLA..293..228B, PhysRevD.65.042001, corbitt:021802,
  sheard:051801, miyakawa:022001}, can play a dominant role in the
dynamics.

A stable optical trap for a macroscopic mirror has been presented in
Ref.~\onlinecite{corbitt:150802}, exploiting the radiation pressure of
two laser fields detuned from cavity resonance to create
simultaneously an optical spring and an optical damping force.  When
mechanical forces coupling the mirror with the outside world are
negligible in comparison with optical forces, this system becomes
nearly immune to the deleterious interaction with its thermal
environment.  This makes it a promising candidate to exhibit quantum
effects including entanglement, a prospect to be evaluated here.

\begin{figure}
  \begin{center}
    \includegraphics[width=8cm]{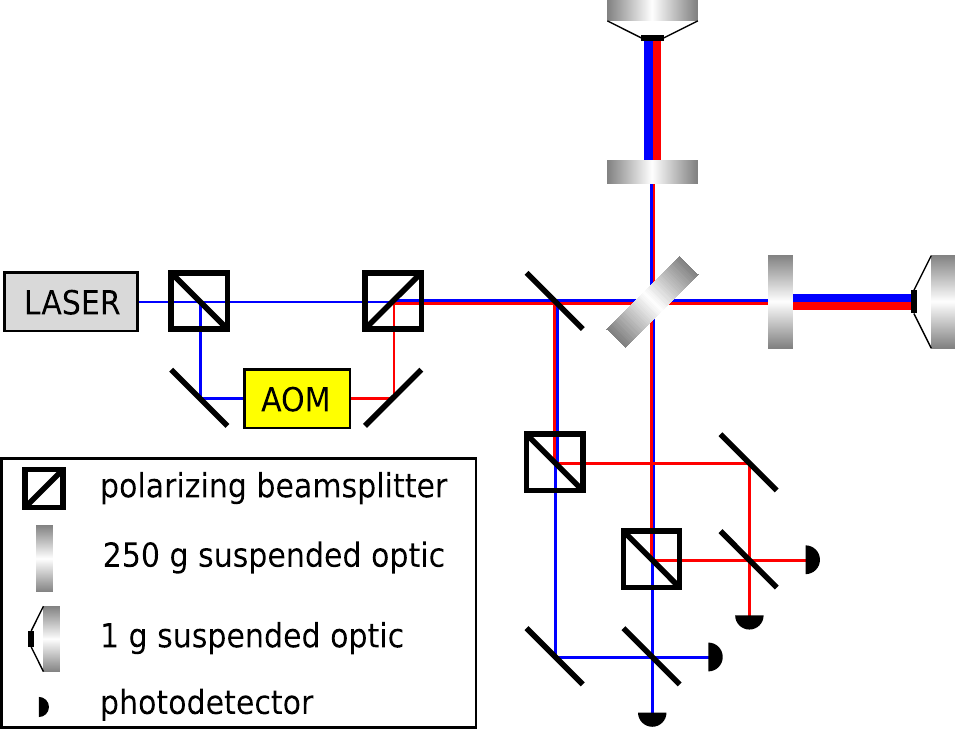}
    \caption{\label{fig:sch} Schematic of an optical trapping and
      homodyne readout apparatus for the differential mode of a
      Fabry-Perot Michelson interferometer.  Each arm cavity comprises
      a highly reflective, low-mass end mirror and a massive input
      mirror of finite transmissivity.  The system is driven by two
      orthogonally polarized laser beams: a strong ``carrier'' field,
      and a weaker frequency-shifted ``subcarrier'' created by an
      acousto-optic modulator (AOM).  Each optical field is monitored
      using a balanced homodyne readout.  Feedback loops required to
      hold the interferometer on resonance are not shown.}
  \end{center}
\end{figure}

\textit{Entanglement criterion.}---It is known that entanglement of a
bipartite continuous-variable system can be recognized by inspecting
its variance matrix for evidence of non-classical correlation.  This
$4 \times 4$ symmetric matrix contains the second order moments
between elements of a vector of observables $\textbf{u} = [Q_1,
\allowbreak P_1, \allowbreak Q_2, \allowbreak P_2]^T$ (i.e., the
canonical positions $Q_j$ and momenta $P_j$ of subsystems $j \in
\{1,2\}$), and is defined as follows:
\begin{equation}
  \label{eq:varmatrix}
  \boldsymbol{V} = \left[
    \begin{matrix}
      \boldsymbol{V}_{\!11} & \boldsymbol{V}_{\!12} \\
      \boldsymbol{V}_{\!12}^T & \boldsymbol{V}_{\!22}
    \end{matrix}
  \right] ; \quad
  \boldsymbol{V}_{\!jk} = \left[
    \begin{matrix}
      \langle Q_j Q_k \rangle_+ & \langle Q_j P_k \rangle_+ \\
      \langle P_j Q_k \rangle_+ & \langle P_j P_k \rangle_+
    \end{matrix}
  \right] .
\end{equation}
Here $\textbf{u}$ is assumed to have zero mean (the steady-state value
$\bar{u}_j$ of each element has been subtracted, leaving only
fluctuating terms).  The quantity $\langle u v \rangle_+$ denotes the
symmetrized average $\langle uv + vu \rangle / 2$.

The Peres-Horodecki entanglement criterion~\cite{PhysRevLett.77.1413},
as stated for continuous-variable systems by
Simon~\cite{PhysRevLett.84.2726}, establishes that the system is
entangled whenever the time reversal of one subsystem only (e.g.\ $P_2
\rightarrow -P_2$) would result in a variance matrix that no longer
satisfies the uncertainty principle.  Stated mathematically,
separability constrains the variance matrix by requiring $4 \det
\boldsymbol{V} > \Sigma - \frac{1}{4}$, where $\Sigma = \det
\boldsymbol{V}_{\!11} + \det \boldsymbol{V}_{\!22} - 2 \det
\boldsymbol{V}_{\!12}$.  Further, one may define the logarithmic
negativity \cite{plenio:090503} in terms of $\boldsymbol{V}$:
\begin{equation}
  \label{eq:neg}
  E_N = \max\left[0, -\frac{1}{2}\ln\left(2\Sigma - 2\sqrt{\Sigma^2 - 4 \det
        \boldsymbol{V}}\right)\right] .
\end{equation}
This entanglement measure quantifies the degree to which the
Peres-Horodecki criterion has been violated~\cite{
  1751-8121-40-28-S01}.  Note that the preceding statements presume
(dimensionless) canonical commutation relations between the elements
of $\textbf{u}$: $[Q_j,P_k] = i \delta_{jk}$, and $[Q_j,Q_k] =
[P_j,P_k] = 0$.

\textit{Opto-mechanical dynamics.}---A schematic of a trapped mirror
system is shown in Fig.~\ref{fig:sch}.  To compute its second order
moments, we first write down its linearized, Heisenberg-picture
equations of motion, which are derived using the quantum Langevin
approach (cf.~\cite{GZQuantNoise,
  2003qic-gmt-information-theoretic-ponderomotive,
  paternostro:250401}).  They can be expressed in the succinct form
\begin{equation}
  \label{eq:lqle}
  \dot{\textbf{u}}_{\text{ic}} =
  \boldsymbol{K}\textbf{u}_{\text{ic}} + \textbf{u}_{\text{in}} .
\end{equation}
This operator equation relates a vector of intra-cavity coordinates,
$\textbf{u}_{\text{ic}} = [q, \allowbreak p, \allowbreak X_1,
\allowbreak Y_1, \allowbreak X_2, \allowbreak Y_2]^T$, and a vector of
input noises driving the system, $\textbf{u}_{\text{in}} = [0,
\allowbreak F_{\text{th}}, \allowbreak \sqrt{2\gamma_c}
X_{\text{in},1}, \allowbreak \sqrt{2\gamma_c} Y_{\text{in},1},
\allowbreak \sqrt{2\gamma_c} X_{\text{in},2}, \allowbreak
\sqrt{2\gamma_c} Y_{\text{in},2}]^T$, which arise from coupling with
the environment.  Elements of $\textbf{u}_{\text{ic}}$ include the
coordinates $q, p$ of the mirror, and the cavity optical mode
quadrature operators defined by $X = (a^\dagger + a)/\sqrt{2}, Y =
i(a^\dagger - a)/\sqrt{2}$.  Elements of $\textbf{u}_{\text{in}}$
include a Langevin force $F_{\text{th}}$ driving Brownian motion of
the mirror, and the vacuum noises $X_{\text{in},j}, Y_{\text{in},j}$
entering each cavity mode.  The coupling matrix is
\begin{equation}
  \label{eq:K}
  \boldsymbol{K} =
  \left[
    \begin{matrix}
      0 & 1/m & 0 & 0 & 0 & 0 \\
      -m\omega_m^2 & -\gamma_m & \hbar G_1 & 0 & \hbar G_2 & 0 \\
      0 & 0 & -\gamma_c & \Delta_1 & 0 & 0 \\
      G_1 & 0 & -\Delta_1 & -\gamma_c & 0 & 0 \\
      0 & 0 & 0 & 0 & -\gamma_c & \Delta_2 \\
      G_2 & 0 & 0 & 0 & -\Delta_2 & -\gamma_c
    \end{matrix}
  \right] .
\end{equation}
Here the cavity mode operators are represented in the frame rotating
with their drive fields, so that only their detunings appear in the
equations, and $G_j = \alpha_j \* \omega_c/L$ parametrizes the
opto-mechanical coupling.  The intra-cavity amplitude near resonance
is related to the incident power $I_j$ by $\alpha_j^2 = 4 I_j \gamma_c
/ [\hbar\omega_c(\gamma_c^2 + \Delta_j^2)]$, and the detuning of each
field is $\Delta_j = (1-\bar{q}/L)\omega_c - \omega_j$.  All other
parameters are defined in Table~\ref{t:params}.

\begin{table}
  \caption{\label{t:params} Parameters and their nominal values.}
  \begin{ruledtabular}
    \begin{tabular}{lcc}
      mirror resonant frequency & $\omega_m/2\pi$ & 1 Hz \\
      mirror damping rate & $\gamma_m/2\pi$ & 1 $\mu$Hz \\
      mirror reduced mass & $m$ & 0.5 g \\
      cavity resonant frequency & $\omega_c/2\pi$ & $c/(1064\text{
        nm})$ \\
      cavity linewidth (HWHM) & $\gamma_c/2\pi$ & 9.5 kHz \\
      cavity length & $L$ & 1 m \\
      carrier power & $I_1$ & 5 W \\
      carrier detuning & $\Delta_1$ & $-3 \gamma_c$ \\
      subcarrier power & $I_2$ & 0.3 W \\
      subcarrier detuning & $\Delta_2$ & $\gamma_c/2$ \\
      ambient temperature & $T$ & 300 K
    \end{tabular}
  \end{ruledtabular}
\end{table}

Taking the Fourier transform ${\mathcal F}\{f(t)\} =
(2\pi)^{-1/2}\*\int dt\,f(t) \* e^{-i \Omega t}$, it is
straightforward to solve Eq.~\ref{eq:lqle} algebraically for
$\textbf{u}_{\text{ic}}$ in terms of
$\textbf{u}_{\text{in}}$~\cite{2003qic-gmt-information-theoretic-ponderomotive}.
To gain insight into the solution, we begin with the case where $G_1 =
G_2 = 0$, decoupling the subsystems.  Then the mirror's equation of
motion is that of a thermally driven pendulum, $q(\Omega) =
\chi_m(\Omega) F_{\text{th}}(\Omega)$, where the mechanical
susceptibility to force is given by $\chi_m(\Omega) = [m(\omega_m^2 +
i\gamma_m\Omega - \Omega^2)]^{-1}$.

Turning on the interaction has two effects on the mirror.  First, it
introduces new driving terms due to radiation pressure noise.  Second,
it alters the mirror's response function.  When motion is slow on the
cavity timescale ($\Omega \ll \gamma_c$), the opto-mechanical
susceptibility may still be written in the form
$\chi_{\text{eff}}(\Omega) = \left[ m (\omega_{\text{eff}}^2 +
  i\gamma_{\text{eff}}\Omega - \Omega^2)\right]^{-1}$, but the
system's new resonance parameters are:
\begin{equation}
  \label{eq:eff}
  \begin{gathered}
    \omega_{\text{eff}}^2 = \omega_m^2 + \sum_j \omega_{\text{eff},j}^2;
    \quad \omega_{\text{eff},j}^2 = -\frac{\hbar G_j^2
      \Delta_j}{m(\gamma_c^2 + \Delta_j^2)} \\
    \gamma_{\text{eff}} = \gamma_m + \sum_j \gamma_{\text{eff},j}; \quad
    \gamma_{\text{eff},j} = -\frac{2\gamma_c
      \omega_{\text{eff},j}^2}{\gamma_c^2+\Delta_j^2}
  \end{gathered}.
\end{equation}

The coupling strengths and detunings of the two optical fields can be
chosen so that the effective resonant frequency and damping rate have
positive sign, and are dominated by terms of optical origin.  These
are the conditions needed to realize a stable optical
trap~\cite{corbitt:150802}.

\textit{Output variances.}---The optical fields exiting the cavity are
potentially quantum-correlated, due to the coupling of their
intra-cavity amplitude and phase with the motion of a common mirror.
To study these correlations, the variance matrix of the output fields
is obtained from the solution to Eq.~\ref{eq:lqle} via the cavity
input-output relation, $a_{\text{in}} + a_{\text{out}} =
\sqrt{2\gamma_c} a_{\text{ic}}$.  First, as $i\Omega$ occurs
asymmetrically in the frequency-domain equations, the operators must
be made Hermitian by combining the positive and negative frequency
parts: $O^H(\Omega) = (O(\Omega) + O(-\Omega))/\sqrt{2}$.
Subsequently, one finds the variance matrix of the output spatial mode
at sideband frequency $\Omega$, in terms of the correlation spectra of
the noise inputs, which are~\cite{GZQuantNoise}:
\begin{equation}
  \label{eq:inputcov}
  \begin{gathered}
    \langle F_{\text{th}}(\Omega) F_{\text{th}}(\Omega^\prime) \rangle
    = 2\gamma_m m \hbar\Omega N(\Omega)\delta(\Omega + \Omega^\prime) \\
    \langle a_{\text{in},j}(\Omega)
    a^\dagger_{\text{in},k}(\Omega^\prime) \rangle =
    \frac{1}{2}\delta_{j,k}\delta(\Omega + \Omega^\prime) \\
    \langle a_{\text{in},j}(\Omega) a_{\text{in},k}(\Omega^\prime)
    \rangle = \langle a^\dagger_{\text{in},j}(\Omega)
    a^\dagger_{\text{in},k}(\Omega^\prime) \rangle = 0
  \end{gathered}
\end{equation}
with $N(\Omega) = (e^{\hbar\Omega/k_B T} - 1)^{-1}$.  Moreover, the
only non-vanishing commutator among the output fields is
$[X^H_{\text{out},j}(\Omega),\allowbreak
Y^H_{\text{out},j}(\Omega^\prime)] = i \delta(\Omega-\Omega^\prime)$.

Applying Eq.~\ref{eq:neg} to these modes, one can show that for
$\Omega \ll \omega_{\text{eff}}$, the logarithmic negativity of the
output fields is approximately constant and can be written simply:
\begin{equation}
  \label{eq:outneg}
  E_{N,\text{out}} = -\frac{1}{2}\ln\left(1+2\xi\left[\Theta -
      \sqrt{\Theta^2 + \xi^{-1}}\right]\right),
\end{equation}
where $\xi$ and $\Theta$ are dimensionless quantities parametrizing
the entangler strength, and the degradation due to thermal noise,
respectively.  They are defined as:
\begin{equation}
  \label{eq:outnegparm}
  \begin{gathered}
    \xi = \frac{4 \gamma_c^2}{\Delta_1 \Delta_2}
    \frac{\omega_{\text{eff},1}^2
      \omega_{\text{eff},2}^2}{\omega_{\text{eff}}^4} \\
    \Theta = 1 - \frac{\gamma_m}{2\gamma_c} \frac{k_B T}{\hbar
      (\Delta_1/\omega_{\text{eff},1}^2 +
      \Delta_2/\omega_{\text{eff},2}^2)^{-1}}
  \end{gathered}.
\end{equation}

\textit{Experimental prospects.}---An experiment must contend with
technical noise sources such as seismic and laser noise, as well as
the fundamental noises (vacuum and suspension thermal) that are
included in the treatment given here.  A detailed noise study exists
for the case where light sensing the motion of two gram-scale mirrors
is optically recombined in a Fabry-Perot Michelson
interferometer~\cite{corbitt:023801}.  A schematic description of a
proposed experiment is shown in Fig.~\ref{fig:sch}, and relevant
parameters are summarized in Table~\ref{t:params}.  In such a
configuration, the differential motion degree of freedom may be
treated as a single cavity wherein common mode laser technical noise
largely cancels.  In addition, strong restoring and damping forces are
supplied to the suspended mirrors by radiation pressure of two detuned
optical fields.  Consequently the resonant frequency is shifted by 3
orders of magnitude, from $\omega_m/2\pi = 1$~Hz to
$\omega_{\text{eff}}/2\pi \approx 2.3$~kHz, with no concomitant
increase in the mechanical coupling to the environment.  The mirror's
response to all external force noises at frequencies well below
$\omega_{\text{eff}}/2\pi$ is thereby suppressed by the factor
$\omega_m^2/\omega_{\text{eff}}^2$.  This combination of noise
cancellation and suppression should expose the fundamental noises in a
frequency band below $\Omega/2\pi \sim 1\text{ kHz}$.  Within this
spectral window, prospects for entanglement can be evaluated using the
analysis described above.

\begin{figure}
  \begin{center}
    \includegraphics[width=8cm]{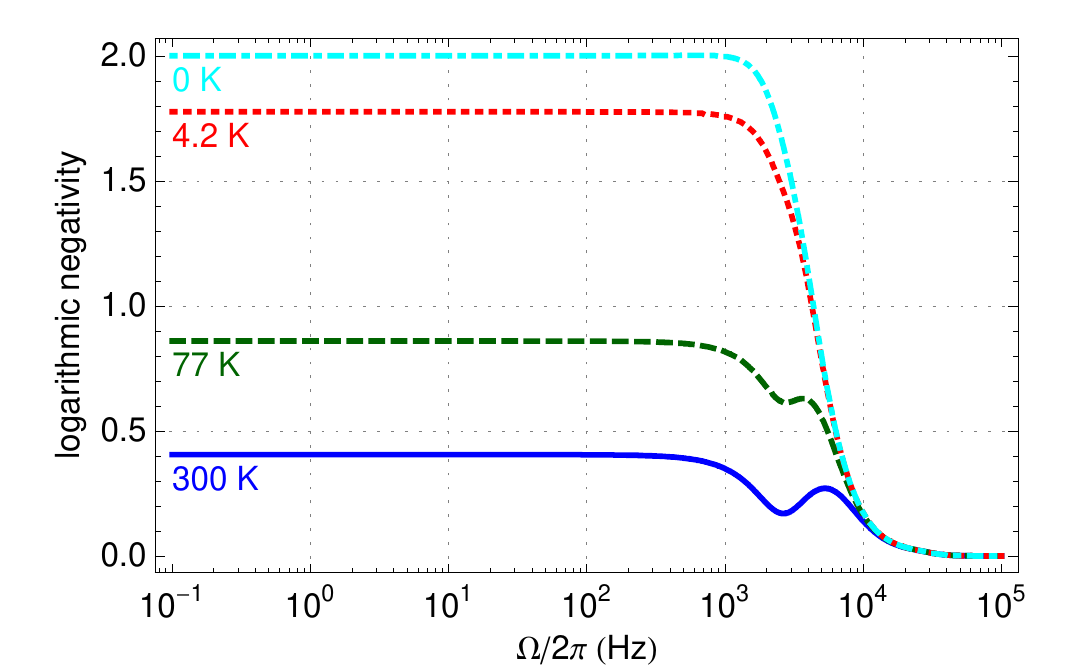}
    \caption{\label{fig:neg} Predicted logarithmic negativity spectra
      for entanglement of the output carrier and subcarrier fields,
      plotted for various ambient temperatures $T$.  Additional
      parameters are specified in Table~\ref{t:params}.}
  \end{center}
\end{figure}

Results of numerical evaluation of $E_{N,\text{out}}(\Omega)$ are
presented in Fig.~\ref{fig:neg}, showing that entanglement of the
output light should be produced within the frequency band of interest,
and that it is remarkably robust against thermal noise --- even
surviving a room-temperature environment.  The spectra are flat until
a thermally-induced depression at the effective resonant frequency
$\omega_{\text{eff}} \approx 2.3$~kHz, with a cut-off at the cavity
linewidth $\gamma_c/2\pi \approx 9.5$~kHz; the magnitude at low
frequency is well approximated by Eq.~\ref{eq:outneg}.

\begin{figure}
  \begin{center}
    \includegraphics[width=8cm]{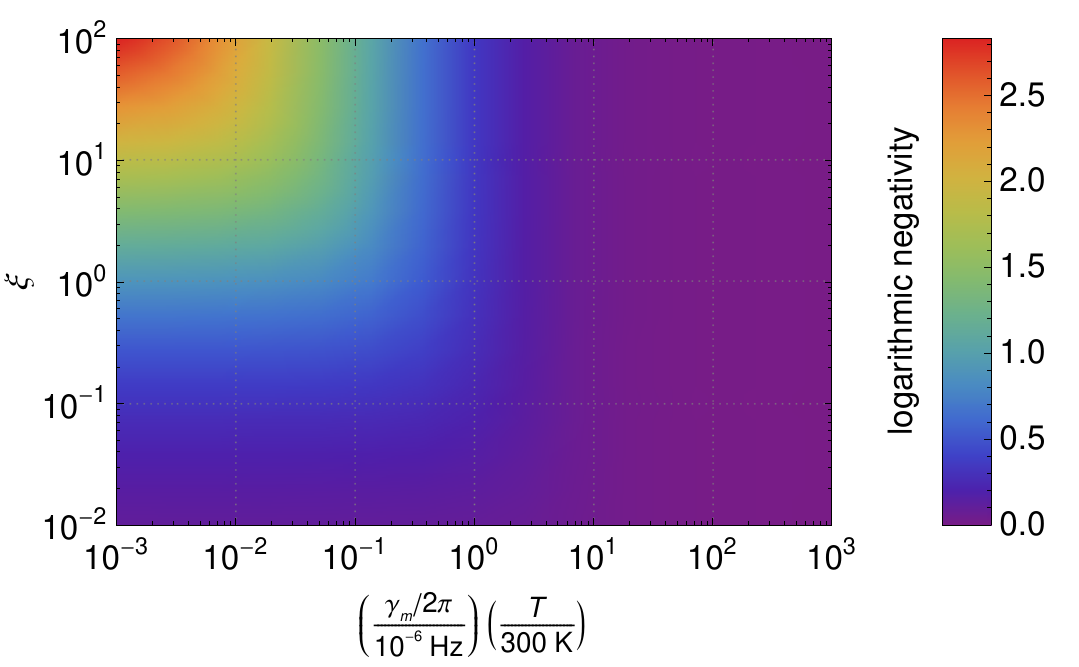}
    \caption{\label{fig:negmap} Logarithmic negativity of output
      carrier-subcarrier entanglement in the frequency-independent
      regime.  The independent variable on the horizontal axis
      corresponds to the thermal noise degradation parameter
      $\Theta-1$.}
  \end{center}
\end{figure}

Given the assumptions of Table~\ref{t:params}, the entangler strength
parameter is $\xi \approx 13.2$, and the thermal degradation parameter
is $\Theta \approx 1.8$ at room temperature.  In this ``strong
entangler'' limit, one finds
\begin{equation}
  \label{eq:strongneg}
  E_{N,\text{out}} \xrightarrow{\xi \gg 1}
  -\frac{1}{2}\ln\left(1 - \frac{1}{\Theta} + \frac{1}{4}
    \frac{\xi^{-1}}{\Theta^3}\right),
\end{equation}
from which it is evident that the magnitude of the negativity is being
constrained solely by $\Theta$, as depicted in Fig.~\ref{fig:negmap}.
Although within the limits of our approximations the output
entanglement never totally vanishes, a soft, low-loss suspension is
necessary to avoid diminution of the logarithmic negativity by thermal
noise.  To capture an appreciable fraction of the available
entanglement, for a suspension with $\omega_m/2\pi \sim 1$~Hz a
quality factor $Q_m = \omega_m/\gamma_m \sim 10^6$ is required.  This
is experimentally challenging but can be achieved, for example, in
suspensions constructed of monolithic fused
silica~\cite{PhysRevLett.85.2442}.

Finally, we remark that homodyne detection of both output optical
fields provides a way to measure their covariance in any desired
quadrature, permitting the entanglement borne by these fields to be
quantified in an experimental setting.  Such techniques have been
demonstrated on entangled light produced by optical parametric
oscillator systems~\cite{2005JOptB...7..577L}.

\begin{figure}
  \begin{center}
    \includegraphics[width=8cm]{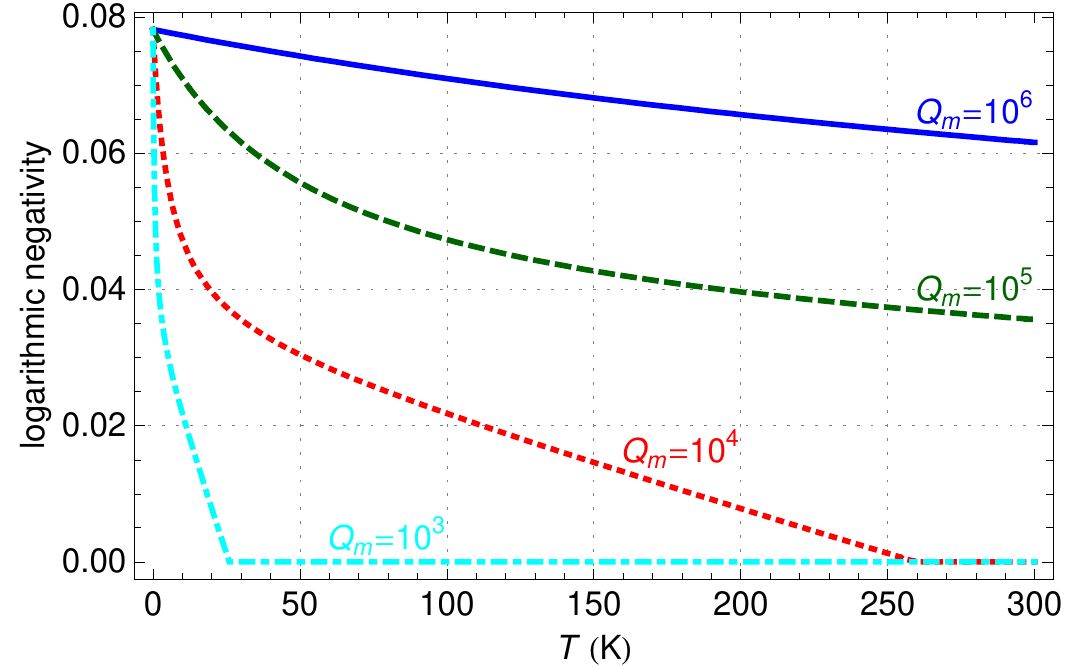}
    \caption{\label{fig:icneg} Logarithmic negativity for bipartite
      entanglement of the mirror with the intra-cavity carrier field.}
  \end{center}
\end{figure}

\textit{Opto-mechanical entanglement.}---The mirror not only generates
the optical entanglement described above, but also can itself become
entangled with the intra-cavity fields.  The intra-cavity variance
matrix can be recovered either from the above analysis via the inverse
Fourier transform, or by applying the Lyapunov equation (as in
Ref.~\onlinecite{PhysRevLett.98.030405}), with correspondence between
the two methods providing a valuable check on the numerics.  The
results, plotted in Fig.~\ref{fig:icneg}, are subject to the caveat
that the noise model considered here is expected to be valid only in a
limited frequency band below 1~kHz, due to the presence of unmodeled
technical noise sources at other frequencies.  We note, however, that
fundamental noise sources do not preclude generating this form of
entanglement of the system.  It has been proposed to verify the
opto-mechanical entanglement with the use of an auxiliary cavity and
homodyne detection on the output light~\cite{PhysRevLett.98.030405}.

\textit{Concluding remarks.}---We have evaluated the capabilities of a
ponderomotive entangler in a novel parameter regime that we believe is
experimentally achievable.  A singular feature of the system under
consideration is the production of entanglement by gram-scale
mechanical objects, while immersed in a room-temperature environment.
Notable attributes of the apparatus that should allow observation of
this entanglement include differential mode noise cancellation in the
Fabry-Perot Michelson interferometer configuration, and the isolation
from external forces supplied by the stiff optical trap.  Construction
and operation of this apparatus are underway at our laboratory.

\begin{acknowledgments}
  We would like to thank our colleagues at the LIGO Laboratory,
  especially Timothy Bodiya and Nicolas Smith; the AEI-Caltech-MIT MQM
  group; and Florian Marquardt of LMU, for helpful discussions.  We
  gratefully acknowledge support from National Science Foundation
  Grants No.~PHY-0107417 and No.~PHY-0457264.
\end{acknowledgments}

\end{document}